\documentstyle[12pt,psfig]{article}
\def\BA{\begin{eqnarray}}
\def\BE{\begin{equation}}
\def\EA{\end{eqnarray}}
\def\EE{\end{equation}}
\def\eps{\varepsilon}

\def\gtsim{\lower-0.45ex\hbox{$>$}\kern-0.77em\lower0.55ex\hbox{$\sim$}}
\def\ltsim{\lower-0.45ex\hbox{$<$}\kern-0.77em\lower0.55ex\hbox{$\sim$}}

\begin{document}
\title{\bf Low $x$ - Scattering as a Critical Phenomenon}

\author{Hans J.~Pirner}
\bigskip
\maketitle
{\it
    Institut f\"ur Theoretische Physik der
    Universit\"at Heidelberg, Germany
  \\
    Max-Planck-Institut f\"ur Kernphysik Heidelberg, Germany
  }
\date{}

\begin{abstract}
We discuss deep inelastic scattering at low $x$  as a critical 
phenomenon in $2+1$  space - time
dimensions.  QCD (SU2) 
near the light cone becomes a critical theory in the limit of 
$\lim x \rightarrow 0$ with a correlation mass 
$m(x) \propto x^{\nu/2}$. 
We conjecture that the perturbative dipole 
wave function of the virtual photon in the region $1/Q<x_{\bot}<1/m$ obeys
correlation scaling $\Psi \propto (x_{\bot})^{-(1+n)}$
before exponentially
decaying for distances larger than the inverse correlation mass. 
This behavior 
combined with an $x$ -independent 
dipole proton cross section gives a longitudinal
structure function which shows the dominant features of the experimental
data. For SU3 QCD a similar second order phase transition is possible 
in the presence of quark zero modes on the light cone.  
\end{abstract}

\section{Introduction}
Perturbative QCD has partially been successful to 
explain low $x$ physics. The BFKL-equation \cite {BFKL}  gives an 
increasing high energy $x$ cross section.  Especially interesting is
the formulation of the BFKL-pomeron with the help of
Wilson lines \cite {Bal}. A perturbative evolution equation of 
Wilson line correlation functions with respect to their  slope
against the light cone 
reproduces the BFKL pomeron. Higher orders \cite{Alt} need
special resummation techniques \cite{ABF} to extract reasonable
information.   
There are also unitarity corrections from the multiple scattering
of the numerous evolved dipoles which may be equally important as  the 
NLO-BFKL-corrections
\cite {Mueller,Bartels}.
In addition  the characteristic transverse momenta in the BFKL equation drift 
to smaller values
making a perturbative calculation questionable.

We follow the very promising
Wilson line method, but   use a nonperturbative lattice approach 
for an effective near light cone Hamiltonian. 
Wilson loops already have been a very fruitful tool to 
calculate the soft photon hadron cross section \cite {DGKP}.
In standard light cone theory a trivial vacuum
is the starting basis \cite{Brodsky}.
Our alternative approach \cite{NPFV,HJP} 
based on the work of refs. \cite {PF,L}
includes  the important nonperturbative
physics  near the light
cone in a solvable form related to the transverse
dynamics of Wilson line operators which 
in modified light cone gauge are functions of 
the $x_-$-independent zero mode
fields
$a_-=A_-(x_\bot)$. 
High energy scattering where the incoming particles propagate near the
light cone and interact mainly through particle exchanges
with transverse momenta is reduced to an effective  $2+1$ dimensional theory.

In this paper we would like to demonstrate 
the relevance of the near light cone formulation of QCD to the
low $x$ limit of structure functions in high energy deep inelastic 
scattering. QCD (SU2) has 
an infrared stable fixed point in the limit when the longitudinal
light like interval determining the boundary conditions 
approaches infinity and the Bjorken variable x goes to zero.
Near this critical point the photon wave function is characterized by two
length scales : the inverse photon virtuality $1/Q$ and 
a correlation length $1/m$. Critical behavior is
expected when the correlation length is larger than the intrinsic
size of the perturbative $q \bar q $ dipole in the photon. 
	By setting up     high energy scattering as a theory of Wilson
lines near the light cone, we are able to determine the low   $x$ behavior 
of the correlation mass $m$ and thereby get a handle on the photon 
wave function
at large distances where perturbative treatments experience infrared
instabilities.

Near light cone QCD has a nontrivial vacuum 
in the light cone limit. 
The zero mode
Hamiltonian depends on an effective coupling constant containing
an additional parameter $\eta$ which labels the coordinate system.
The light cone is approached when the parameter $\eta$ goes to zero.
Generally, if the zero mode theory has massive excitations,
these masses diverge like $\frac{1}{\eta}$
in the light cone limit ${\eta \rightarrow 0}$ and decouple.
Massless excitations in the 
zero mode theory, however,  influence the
light cone limit. Genuine nonperturbative techniques must be
used to investigate their behavior. We claim that the increasing
high energy cross sections are a directly observable consequence
of these massless excitations.  
We choose the following near light cone coordinates which
smoothly interpolate between the Lorentz and light front
coordinates :
\BA
  x^t ~=~ x^{+} &=& \frac1{\sqrt2} 
       \left\{ \left(1 + \frac{\eta^2}{2} \right)
       x^{0} + \left(1 - \frac{\eta^2}{2} \right) x^{3} 
       \right\} ~,\nonumber \\
          x^{-} &=& \frac1{\sqrt2} \left( x^{0}-x^{3} \right)~.
\label{Coor}
\EA
The transverse coordinates $x^1$, $x^2$ are unchanged;
$x^t = x^{+}$ is the new time coordinate, $x^{-}$ is a 
spatial coordinate. As finite quantization volume we 
take a torus with extension $L$ in spatial ``-'' direction, 
as well as in ``1, 2'' 
direction. The scalar product of two 4-vectors $x$ and $y$
is given with $\vec{x}_\bot \vec{y}_\bot = x^1 y^1 + x^2 y^2$ as
\BA
  x_\mu y^\mu & = & x^- y^+ + x^+ y^- - \eta^2 x^- y^-
                - \vec{x}_\bot \vec{y}_\bot \nonumber \\
              & = & x_- y_+ + x_+ y_- + \eta^2 x_+ y_+
                - \vec{x}_\bot \vec{y}_\bot ~.
\label{scalpr}
\EA
Obviously, the light cone is approached as the parameter $\eta$ 
goes to zero.

Now consider high energy photon proton scattering at small $x=Q^2/s$,
where $s=W^2$ is the cm energy squared. Using the photon vector
$q, q^2=-Q^2$ and the proton vector $p,p^2=m^2 \approx 0$ 
we can define two light like vectors 
\BA
  e_1 &=& q- \frac{q^2}{2 pq}p\nonumber \\
  e_2 &=& p.
\EA

For finite energies the vector of the photon $q$ can be calculated as
linear combination of the light like vector $e_1$ with a small amount
of $e_2$ admixed .
\BA
  e_{\eta} &=& q +xp -\frac{\eta^2}{2} p\nonumber \\
           &=& e_1-\frac{\eta^2}{2}e_2.
\EA 

In the limit of infinite energies the mixing $\eta$ related to the 
Bjorken variable $x$ vanishes as $\frac{\eta^2}{2}=x$.
Therefore it is very natural to formulate high energy scattering
in near light cone coordinates. 

The appropriate gauge fixing procedure near the light cone is 
modified light-cone gauge 
$\partial_-A_- = 0$ which allows  zero modes dependent on the transverse 
coordinates. These zero mode fields carry zero linear momentum $p_-$
in near light cone coordinates, but finite amount of $p_0+p_3$.
They correspond to "wee" or low $x$ - partons in the language of 
 Feynman. In colour $SU(2)$ the zero-mode fields $a^3_-(x_\perp)$
can be chosen colour diagonal proportional to $\tau^3$. 
The use of an axial gauge is ideally suited to  the light-cone 
Hamiltonian.
The asymmetry of the background zero mode coincides 
with the asymmetry of the space coordinates on the light cone. 
The zero-mode fields describe disorder fields. Depending on the
effective coupling the zero-mode transverse system is in a
massive or massless phase. 
In a previous paper \cite{HJP} we calculated the transverse 
correlation length of the reduced $2+1$ dimensional theory
as a function of $\eta$ describing the nearness to the light cone.   
Since the  light cone limit 
is synchronized with the continuum limit
a considerable  simplification of the QCD Hamiltonian can be achieved.

\section{Near Light Cone QCD Hamiltonian} 

In ref. \cite{NPFV}  the near light cone Hamiltonian has been derived.
We refer to this paper for further details.
The light cone Hamiltonian on the 
finite light like $x^-$ interval of length L has Wilson line or 
Polyakov operators
similarly to QCD formulated on a finite interval in imaginary time
at finite temperature.
\BE
  P(\vec{x}_\bot) = \frac12 \, \mbox{tr~P} 
  \exp\left(ig \int\limits_0^L dx^-A_-(\vec{x}_\bot,x^-)\tau^3/2\right)\,.
\label{Polyakov}
\EE

Recently the importance of the Wilson line phase operators
for deep inelastic scattering has been demonstrated in the context
of shadowing corrections \cite {BHMPS}. Contrary to naive light cone 
gauge $A_-=0$
the rescattering of the quark in the photon changes the deep inelastic
cross section. 
The   dynamics of these
Polyakov operators is determined by the QCD SU2 Hamiltonian $H$ 
which is a functional of the fields $A_i,a_-$
and the canonical conjugate momenta  $\Pi_i,p^-$, which
represent the electric fields. Here
we give only the gluonic part, we discuss the quarks later in connection
with QCD SU3 :
\BE
  H[\Pi_i,A_i,p^-,a_-] = \int d^3\!x \,{\cal H}(\vec{x})\,,
\label{Ham}
\EE
with
\vskip-7mm
\BA
  {\cal H} & = & \mbox{tr} 
    \left[ \partial_1 A_2 - \partial_2 A_1 - ig [A_1, A_2] \right]^2 
  + \frac1{\eta^2} \,\mbox{tr}
    \left[ \vec\Pi_\bot -
       \left( \partial_ - \vec{A}_\bot - ig [a_-,\vec{A}_\bot] \right)
    \right]^2
    \nonumber \\
  & + &  \frac1{\eta^2} \,\mbox{tr}
    \left[ \frac1L \vec{e}_\bot^{\;3} - \nabla_\bot a_- \right]^2 
    + \frac1{2 L^2} p^{-\dagger}(\vec{x}_\bot) p^{-}(\vec{x}_\bot)
    \nonumber \\
  & + & \frac1{L^{2}} \int\limits_0^L dz^- \int\limits_0^L dy^-
    \sum_{p,q,n}\!' 
      \frac{G_{\bot qp}(\vec{x}_\bot, z^-)%
            G_{\bot pq}(\vec{x}_\bot, y^-)}{\left[
      \frac{2\pi n}{L} + g(a_{-,q}(\vec{x}_\bot)
         - a_{-,p}(\vec{x}_\bot)) \right]^2 }
      e^{ 2\pi i n (z^- - y^-)/L} ~.
\label{Hamil}
\EA
The prime indicates that the summation is restricted to $n \neq 0$ 
if $p = q$. The operator $G_{\perp}$ gives the right hand side of
Gauss' law:
\BE
  G_{\perp}\left(\vec{x}\right) =
     \vec{\nabla}_\perp \vec\Pi_\perp \left(\vec{x}\right)
   + g\varepsilon^{ab3} \frac{\lambda^a}2 \vec{A}\,^{b}_\perp
     \left(\vec{x}\right) \left(\vec\Pi\,^{3}_\perp
     \left(\vec{x}\right)
   - \frac1L \vec{e}\,^{3}_\bot \left(\vec{x}_\perp\right)\right) 
   + g\rho_m(\vec{x}) ~,
\label{Gop}
\EE
with $\rho_m$ the matter density. The above Hamiltonian shows rather 
clearly that a naive limiting procedure $\eta \rightarrow 0$ does not
work. There are severe divergences in this limit. The diverging
terms reappear in the usual light cone Hamiltonian as constraint equations
which are extremely difficult to solve on the quantum level in $3+1$
dimensions. 
The zero-mode part $h(a_-(x_\bot),p^-(x_\bot))$
of the full Hamiltonian $\cal H$
is coupled to the
three-dimensional modes of the Hamiltonian. In the following
we will concentrate on universal properties of the zero-mode Hamiltonian
which survive the renormalization of the $(2+1)$ transverse dynamics
due to the coupling to the $(3+1)$  dimensional residual Hamiltonian.
The zero-mode
Hamiltonian contains  the Jacobian $J(a_-)$  from the Haar measure of 
$SU(2)$
$J\left(a_-(\vec{x}_{\bot})\right) = 
   \sin^2 \left( \frac{gL}{2}a_-(\vec{x}_{\bot})\right)$ .   
The measure stems from gauge fixing and also appears in the functional
integration volume element for calculating matrix elements.
It is convenient to introduce dimensionless variables 
$ \varphi$ which  vary in a compact domain $0 \leq \varphi \leq \pi$
and characterize the Wilson lines $P$.

\BA
  \varphi(\vec{x}_\bot) &=& \frac{gL a_-(\vec{x}_\bot)}{2} \nonumber\\
  P(\vec x_{\bot})      &=&   cos\varphi(\vec{x}_\bot) 
\EA
We 
regularize the zero-mode Hamiltonian by introducing a lattice
with lattice constant $a$ in transverse directions. 
Next we appeal to the 
physics of the infinite momentum frame and factorize the reduced
true energy from the Lorentz boost factor $\gamma=\sqrt2/\eta$
and the cut-off by defining $h_{\rm red}$
\BE
  h = \frac1{2\eta a} h_{\rm red}.
  \label{hdef}
\EE

For small lattice spacing we obtain the reduced Hamiltonian
\BE
  h_{\rm red} = \sum_{\vec{b}}\left\{ -g^2_{\rm eff}
    \frac1J \frac\delta{\delta\varphi(\vec b)}
          J \frac\delta{\delta\varphi(\vec b)}
   + \frac1{g^2_{\rm eff}} \sum_{\vec\eps}
    \left(\varphi(\vec{b})-\varphi(\vec{b}+\vec\eps\,)\right)^2
  \right\} 
\EE
with the effective coupling constant
\BE
 g^2_{\rm eff} = \frac{g^2L\eta}{4a}.
\EE
In the continuum limit of the transverse lattice theory the lattice size
$a$ goes to zero and/or the extension $L$ of the lattice to infinity.
This limit combined with the light cone limit $\eta \rightarrow 0$
leads to an indefinite behavior of the effective coupling constant.
The critical behavior of the zero-mode theory resolves this 
ambiguity.
In ref. \cite {HJP} we have done a Finite Size Scaling (FSS) 
analysis obtaining 
a second order transition as a function of the coupling $g^2_{\rm eff}$
between a phase with massive
excitations at strong coupling and a phase with massless excitations in weak
coupling.  The critical effective coupling is calculated as  
$g^{*2}= 0.17 \pm 0.03$, which is, however, not a universal 
quantity and subject to radiative corrections from the $(3+1)$ 
dimensional modes $\Pi_i,A_i$. 
A calculation in the epsilon expansion \cite{LZ}
gives the zero of the $\beta$-function as an infrared stable fixed point. 
Therefore, the combined  limit of large dimensions $L/a$ and light like
coordinates $\eta \rightarrow 0$ is well defined. 
Using the running coupling constant, we get the
critical exponent for the correlation mass $m$  

\BA
  m &=& m_0 \left(g^2_{\rm eff}-g^{*2}\right)^{\nu} \nonumber\\
    &=& m_0 \left( \frac{g^2 L \eta}{4 a}-g^{*2}\right)^{\nu}
\EA
with $\nu=0.63   $. The critical theory is in the same universality
class as the $3$- dim Ising model. Therefore, we use in the following
the critical exponents of the Ising model.

To match the Hamiltonian lattice calculation with scattering we 
consider a lattice with a lattice constant $a \approx \frac{1}{Q}$ 
and an extension $L$
which is larger than the colour coherence length  
of the $q \bar q $ state in the photon-proton 
cm system by an amount $L_0$ . 
The  photon and proton move on almost light like
trajectories and the coherence length grows with $x$ as:
\BE
  \Delta x_-= \frac {1}{Q \sqrt(x)}\ .
\EE

We demand therefore:
\BE
  L = \frac {1}{Q \sqrt{x}} + L_0 
\EE

Identifying the $x$ and $a$ independent limit of the running coupling
$g_{eff}^2$ as the fixed point coupling $g^{*2}$ one finds that 
the correlation mass decreases with $x$ towards the
critical point:
\BE
m \approx \frac{1}{a}(x/x_0)^{\nu/2} \approx Q(x/x_0)^{\nu/2}.
\EE

Near the critical point the Wilson lines experience long range correlations,
which means that dipoles in the photon wave function are correlated over
distances $1/m$. In the intermediate range where $\frac{1}{Q}<x_\bot<
\frac{1}{m}$ the correlation function of Wilson lines is power behaved:
\BE
  <cos \varphi(x_\bot) cos \varphi(0)> \approx \frac{1}{x_\bot^{1+n}}
\EE
where $n=0.04$ in Ising like systems.
For even larger distances $x_\bot>\frac{1}{m}$ the correlation
function decreases exponentially:
\BE
  <cos \varphi(x_\bot)cos \varphi(0)> \approx e ^{-m x_\bot}.
\EE

In the following we construct an effective photon wave function
which has these length scales built in and calculate the
resulting longitudinal $\gamma$ -proton cross section.

\section { Longitudinal Photon Proton Scattering and Critical Behavior}

The longitudinal photon is ideal to investigate high energy cross
sections of hadronic objects with a variable size. Experimental
longitudinal cross sections are until now less accurate
but they are much better suited to investigate the
complex scattering dynamics of small size objects at 
high energies. Since the
light cone wave function of the $q \bar q$ in the longitudinal 
photon is peaked at equal momentum fractions $z =1/2$ an increasing
virtuality of the photon guarantees a decreasing size of the
perturbative photon dipole wave function. In the course of $x$-evolution 
this wave function will develop many dipoles which in general diffuse
into distance scales beyond the original size $1/Q$. This increase in
parton density and/or  size of the photon wave function is generally 
believed to be the origin of the increasing high energy cross section.
In this work, we do not follow the development of the photon dipole state
in detail, we only give a qualitative description of the
effective photon size as a function of $x$ using the results of the
$2+1$ dimensional critical QCD SU2 theory as a guiding principle.

We parameterize the longitudinal photon probability density as:
\BA
\rho_{\gamma}^L&=&\sum_f 6 \alpha \hat e_f^2 4 Q^2 z^2 (1-z)^2 F(\eps x_\bot)\\
\eps&=&\sqrt{Q^2 z(1-z)}\ . 
\EA

If the correlation mass is larger than the perturbative
momentum scale 
$\eps$, then the function $F(\eps x_\bot)= K_0(\eps x_\bot)^2 $.
This is the uninteresting case for small $x$. 
The photon density is modified only in the small $x$ region 
when $x<x_0$. The parameter $x_0$
is not given  by the critical theory,  we choose 
$x_0=10^{-3}$. Thereby we set the scale 
for the running of the
correlation mass $m$ with the effective momentum  
$\eps= \sqrt{Q^2 z (1-z)}$ and $x$.

\BE
m= \eps (\frac {x}{x_0})^{\nu/2}\ .
\EE

Here $\nu=0.63$ is the Ising critical index determining the
Wilson line correlations.

For 
$x<x_0$ i.e. for $m<\eps$ we replace  the perturbative photon
density using the correlation functions of the critical theory
eqs. (17,18).
\BA
F(\eps x_\bot) &=& K_0(\eps x_\bot)^2 \qquad
\textrm{ for  } x_\bot < \frac {1}{\eps} \\
               &=& \frac{K_0(1)^2}{(\eps x_\bot)^{2+2 n}} \qquad
\textrm{for  }  \frac{1}{\eps} < x_\bot < \frac{1}{m} \\
               &=& K_0(1)^2 (\frac{m}{\eps})^{2+2 n}
e^{-2 m(x_\bot-1/m)} \qquad
\textrm{for  }  x_\bot > \frac{1}{m} 
\EA

The above parametrization extends the photon density 
into the scaling region using the scaling index $n=0.04$.
For distances beyond the scaling region the density decays 
exponentially. The connections are made in such a way that the density is
continuous at the respective boundaries of each region. 

The photon density combined with an energy independent 
dipole-proton cross section determines the 
longitudinal structure function $F_L$ and the photon- proton 
cross section:

\BA 
F_L(x,Q^2) &=& \frac{Q^2}{4 \pi^2 \alpha }\sigma_{\gamma p}^{L,tot}\\
\sigma_{\gamma p}^{L,tot}&=&\int d^2 x_\bot \int _0 ^1 dz 
\rho_{\gamma}^L(x_\bot,z) \sigma_{dip}(x_\bot) 
\EA

where we take for definiteness the Golec-Biernat-W\"usthoff \cite {Golec} 
dipole cross section at fixed $x=x_0$

\BA
\sigma_{dip}(x_\bot) &=& 23 [mb] (1-e^{-\frac{x_\bot^2}{4 R_0^2}})\\
        R_0^2(x_0)   &=& \frac{1}{1 GeV^2}
(\frac{x_0}{3*10^{-4}})^{0.29}\ .
\EA

The numerical values entering the above formulas are taken over from
the original reference. Note,  the value $R_0=0.24 fm$ is independent of $x$.

\begin{figure}[ht]
\centerline{\psfig{figure=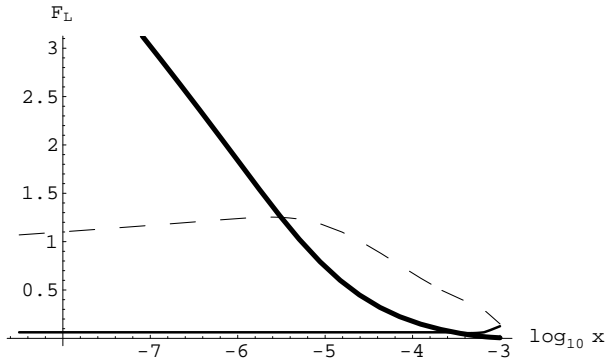,width=8cm}}
\caption{Contributions to the longitudinal structure function 
$F_L(x,Q^2=12 GeV^2)$ from the $x_\bot$ integration over the 
perturbative region $[0,1/\eps]$ (thin line),
the scaling region  $[1/\eps,<1/m]$ (thick line)
and the exponentially decaying photon density  $[1/m,\infty]$ (dashed line)
as a function of $log_{10}(x)$}
\label{cont} 
\end{figure}
Let us  for simplicity first consider the photon wave function at a fixed
virtuality e.g. $Q^2=12  GeV^2$ and fixed 
momentum fraction of the quark and antiquark $\frac{1}{2}=z=(1-z)$.
The respective contributions to $F_L$ as a function of $x$ are
shown in fig.1. A tiny (see thin line) part of $F_L$ comes 
from the very short range part of the photon (eq. 22). 
The
scaling region in the photon density (see thick line in fig.1 and eq. 23 )
leads to a  structure function which behaves as 
$F_L \propto  x^{\nu}$, but is subleading in the interval 
$x_0>x>10^{-5}$. Scaling of this contribution to the 
structure function $F_L$ 
at small $x$ is a consequence of the scaling photon 
wave function depending only on $\eps x_\bot$ and a dipole cross section
$\propto r^2$ for small distances.
With decreasing $x$ one traverses this scaling 
region finally having strong non scaling physics at smaller $x$.
The region of the photon density 
with the exponential decay (cf. dashed curve in fig.1 and eq. 24)
dominates the total cross section until $x \approx 10^{-5}$ where the
correlation length has increased to about $\frac {1}{m} \approx 3 R_0$.
A saturating  
cross section is building up which asymptotically for $x<10^{-5}$  
decreases as $\sigma^{tot} \propto x^{\nu n}$ . This contribution
to the  longitudinal
structure function is approximately proportional to 
$Q^2$, i.e. it does not scale. 

\begin{figure}[ht]
\centerline{\psfig{figure=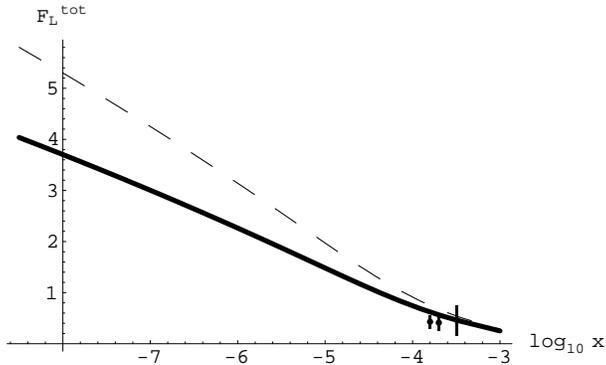,width=8cm}}
\caption{Total longitudinal structure function $F_L^{tot}
(x,Q^2=12 GeV^2)$ calculated
with fixed $z=1/2$ momentum fraction of the quark (dashed curve) and with
integration over $z$-momentum fraction (thick line). The data with error
bars are from the H1-experiment \cite {H1}}
\label{total} 
\end{figure}

In fig. 2 we present the total 
longitudinal  structure function as a function of $x$ for $Q^2=12 
GeV^2$. One sees that effective power of the $x$ - increase in the
currently accessible region is smaller than the asymptotic
critical index $\nu$. In the region $10^{-5.5}<x<10^{-3.5}$ the
effective power is about $0.33$ dominated by the saturating dipole
cross section, which is scanned  by the exponential tail of the
photon wave function. In the next section we discuss the
realistic case taking into account the integration over quark longitudinal 
momenta, which reduces the effective power  mixing length scales
of different $\eps$ for the same photon virtuality $Q$.

An interpretation of low $x$ scattering 
using the  language of the gas liquid transition 
goes as follows: At moderate $x$ the partons are  in a gas phase the
density of which increases with decreasing x. The increasing 
parton density exhibits large fluctuations which produce
critical opalescence at the parton gas-liquid phase transition where 
an infinite correlation length develops. 
Because of the QCD Hamiltonian cf.eq. (11)
one can directly see how this critical behavior  arises from  
gluon dynamics.
At large effective coupling $g^2_{\rm eff} = \frac{g^2L\eta}{4a}$ the 
correlation mass is large, few finite sites with excited
electric fields $p^-(x_\bot)$ exist. The 
dynamics is dominated by a large excitation energy for the half rotors 
$0<\varphi(x_{\bot})<\pi$ .
With decreasing coupling when the  system approaches the
light cone these rotors become strongly coupled in transverse space
and massless excitations develop around aligned 
configurations.


\section{Can Critical Dynamics Be Observed?}

In a theory of total cross sections the nonperturbative 
infrared dynamics in the transverse plane is essential. 
In this section we will discuss three obstacles which
may make it difficult to observe critical behavior in
high energy scattering:
 
\begin {itemize}
\item SU3 dynamics 
\item $z$- smearing of the photon wave function  
\item unitarity effects versus transverse growth

\end {itemize}

It is important
to question whether real QCD with $N_c=3$ really can lead
to  critical behavior. We know from comparable finite 
temperature studies that the critical behavior depends on the number of
colours. SU3 pure glue QCD is in the same universality class as the $Z(3)$
Potts model
and generates a $1$-st order phase transition. The same is expected
for the Wilson line zero mode theory, which would contain now 
two diagonal zero mode fields $a_-^3$ and $a_-^8$ after  Abelian
gauge fixing. As shown in ref. \cite {NPFV}, the full Hamiltonian 
contains a piece where the the zero mode field is coupled to the
negative energy quark field $\psi_-$

\BE
H_{coupling}=-i\frac{2}{\eta^2}\psi^\dagger_-(\partial_--i g
a_-)\psi_-\ .
\EE

The coupling term can act as an ``external magnetic field'' 
in the language of spin models for the
$a_-$ dynamics. A possible quark zero mode $ \psi^\dagger_-\psi_-$ 
would represent the magnetic field for the Potts spin. 
If such a zero mode  exists,
the second order phase transition in SU2 will become a pseudo-critical 
cross over  and a first order transition point in SU3
goes over in a first order
transition line in the coupling constant-quark fermion zero mode  
plane. This line 
ends at a second order end point, where similar critical dynamics as 
described above occurs. One must admit that the existence of such a
coloured zero mode in  $ \psi^{\dagger 3}_-\psi_-^3$  and  
$\psi^{\dagger 8}_-\psi_-^8$  is puzzling and more theoretical 
work is needed to study
its possibility. If there are zero mode coloured gluon fields 
on the light cone, 
it is not excluded  that also the negative energy states 
have zero modes
on light like trajectories, comparable to the filled Dirac sea in the
normal vacuum.

The second question about the averaging of distance scales
through quark $z$-motion can be easily solved by numerical computation. 
In the
full calculation of the structure function the integration over $z$
smears the clear transverse boundaries in $x_\bot$ space 
of the effective photon wave function c.f.eqs. (21-23).
In fig.2 we compare the fully $z$-integrated structure function 
$F_L(x,Q^2=12 GeV^2)$
with the approximate one taken at fixed  $z=1/2$. One sees
the expected flattening of the x- behavior , in the interval $10^{-5.5}<x<
10^{-3.5}$ the  power decreases from $1/x^{0.33}$ to $1/x^{0.29}$.  
For larger $Q^2=45 GeV^2$ the effective power increases to $1/x^{0.45}$ and
$1/x^{0.41}$, respectively. An even smaller value of $x$ decreases the 
power further, because the scaling part of the wave function 
giving  the dominant part of the structure function at very
small x multiplies a saturating dipole proton cross section
and produces therefore a small $x$- decrease. Only at moderate $10^{-5.5}<x<
10^{-3.5}$ when the scaling part of the wave function overlaps with the
$r^2$ dependent dipole cross section, the critical index $\nu$
could be observed in this  part of the cross section increasing $\propto 
(\frac{\eps}{m})^{2-2 n} \propto 1/x^{\nu(1-n)}$.
Unfortunately this part of the cross section from the integration
over $[1/\eps,1/m]$ is smaller than 
the dipole cross section integrated over the interval 
$[1/m,\infty]$. In fig.2 the data points from ref. \cite {H1},
illustrate how far the theory extrapolates beyond the
currently accessible experimental region.

This finding leads us to the third question
about unitarity effects. The numerical calculation
clearly shows that 
saturation effects in the dipole proton  cross
section tame the growth of the photon proton cross section due
to an increasing effective photon 
size at very small $x$. It has been conjectured that the saturating
dipole cross section in the above parametrization is due  to
unitarity effects. For a final proof of such a hypothesis it is, however,
necessary to study the dependence of dipole proton scattering on the
impact parameter. Only the S-matrix $S(b)$ can teach us about unitarity.
We refer to such a study in the context of the stochastic vacuum model 
in ref. \cite {DSSP}. In this work saturation effects are clearly
seen in proton-proton scattering at $W=2.5 TeV$, but no indication in
photon-proton scattering at the same energy is visible. 
Preliminary phenomenological analysis of the data in vector meson
production also do not show big effects \cite {Munier}.

	Clearly on the theoretical side it is necessary 
to take into account the target 
at $x^-=0$ in a calculation with the dynamical gluon fields
$A_\bot(x_-,x_\bot)$. Recently important progress has been made using the
assumption that the low $x$ problem can be studied in perturbation
theory with an effective theory of Wilson lines. The lattice studies
can in principle also treat the problem with an external source at
$x^-=0$ with a given   structure (e.g. dipole state or 
``spin-glass'' \cite {Iancu}) 
in transverse space. Ultimately they always will be the best 
tool to handle the
infrared diffusion of the low $x$ wave function. 
	In this paper we have made a first step extending the
picture of a dilute parton gas into a dense liquid like phase 
at small $x$. Starting from the QCD Hamiltonian
a critically diverging correlation length in transverse space
leads to increasing cross sections in qualitative 
agreement with the observed growth for high  $Q^2$ 
deep inelastic scattering. 
\section*{Acknowledgments}
We thank  H.G. Dosch and G.P. 
Korchemsky for the encouragement 
to complete this work and are grateful to C. Ewerz, M. Ilgenfritz,
B. Kopeliovich, A. Mueller and E.
Prokhvatilov for helpful discussions.

This work has been partially funded through the European TMR Contract
$HPRN-CT-2000-00130$ :Electron Scattering off Confined Partons.


\end{document}